\documentclass[twocolumn]{autart}    

\usepackage{graphicx}          
\usepackage{xcolor,epstopdf}
\usepackage{amsmath,amsfonts,amssymb,mathtools}
\usepackage{enumitem}
\usepackage{booktabs}
\newcommand{\ra}[1]{\renewcommand\arraystretch{#1}}

\newtheorem{problem}{Problem}
\newtheorem{remark}{Remark}
\newtheorem{proposition}{Proposition}
\newtheorem{theorem}{Theorem}
\newtheorem{corollary}{Corollary}
\newtheorem{lemma}{Lemma}
\newtheorem{definition}{Definition}

\newenvironment{proof}{%
    \begin{pf}%
}{%
    \end{pf}%
    \ignorespacesafterend%
}

\newcommand{\isdef}{\stackrel{\triangle}{=}}

\begin{document}

\begin{frontmatter}

\title{Geometrization of Higher-Order Linear Control Laws for Attitude Control on $\mathrm{SO(3)}$} 


\author[First]{Farooq Aslam}\ead{farooq.aslam87@gmail.com},
\author[First,Second]{Hafiz Zeeshan Iqbal Khan}\ead{zeeshaniqbalkhan@hotmail.com},
\author[First]{Muhammad Farooq Haydar}, 
\author[First]{Suhail Akhtar}, 
\author[First]{Jamshed Riaz} 

\address[First]{Institute of Space Technology, Islamabad, Pakistan}
\address[Second]{Centers of Excellence in Science and Applied Technologies, Islamabad, Pakistan}


\begin{abstract}                          
This paper presents a theoretical framework for analyzing the stability of higher-order geometric nonlinear control laws for attitude control on the Special Orthogonal Group $\mathrm{SO(3)}$. In particular, the paper extends existing results on the analysis of PID-type geometric nonlinear control laws to more general higher-order dynamic state-feedback compensators on $\mathrm{SO(3)}$. The candidate Lyapunov function is motivated by quadratic Lyapunov functions of the form $V(x)=x^{\top}Px$ typically considered in the analysis of linear time-invariant (LTI) systems. The stability analysis is carried out in two steps. In the first step, a sufficient condition is obtained for the positive definiteness of the candidate Lyapunov function, and a necessary and sufficient condition for the negative definiteness of the corresponding Lyapunov rate. \textcolor{black}{These conditions ensure 
that the desired equilibrium is almost globally asymptotically stable (AGAS).} 
In the second step, \textcolor{black}{a convex relaxation of the proposed conditions is used} to obtain sufficient conditions in the form of linear matrix inequalities (LMIs). Overall, the approach is motivated  by the widespread use of LMI-based analysis and design tools for LTI systems. \textcolor{black}{To reduce conservatism}, matrix gains are considered for the controller gains as well as the Lyapunov function coefficients.
The applicability of the approach to practical problems is illustrated by designing and analyzing a 21-state geometric nonlinear attitude control law for a multicopter.
\end{abstract}

\end{frontmatter}

\section{Introduction}

Given the conceptual advantages of the rotation matrix parameterization over other attitude representations, several researchers have sought to develop singularity-free and unwinding-free attitude control laws directly on the Special Orthogonal Group $\mathrm{SO(3)}$ \cite{chaturvedi2011rigid}.
%
%
%
%
Among the proposed solutions, geometric nonlinear PID controllers have received considerable attention as they provide a relatively simple and effective way of performing attitude control (see \cite{bullo1995proportional,lee2010geometric,goodarzi2013geometric,maithripala2015intrinsic} and the references therein).
In general, geometric nonlinear PID controllers use proportional and derivative terms to correct the rotational and angular velocity errors, respectively, and an integral term that accounts for fixed or slowly-varying disturbances.
Using these corrections, PID controllers are able to almost globally asymptotically stabilize the desired equilibrium of the closed-loop error system \cite{goodarzi2013geometric}.
In practice, controller gains are often tuned using a combination of linearization-based design methods and $\mathrm{SO(3)}$-based analysis methods \cite{invernizzi2020robust,aslam2023CDC}.
The resulting PID controllers typically possess local performance guarantees as well as almost global asymptotic stability (AGAS) guarantees.

\textcolor{black}{More generally, PID controllers are dynamic compensators which use
a combination of static state-feedback and dynamically evolving states, typically the integrator state.
A case in point is the geometric nonlinear PID-type attitude controller proposed in \cite{goodarzi2013geometric}. This controller, and others like it, may be viewed as a particular instance of a broader class of feedback controllers, namely, higher-order dynamic compensators on $\mathrm{SO(3)}$.
%
This class of compensators also includes lead/lag filters. These filters are often used in feedback control systems as \textit{implementable} PID controllers, with the lead filter acting as the PD controller, and the lag filter as the PI controller.
%
%
%
In addition to PID controllers, performance and hardware implementation requirements sometimes necessitate the use of additional compensators. 
For practical control systems, practitioners often resort to linearization-based design methods for designing the various filters and compensators \cite{invernizzi2020robust}.
However, 
linearization-based methods typically yield controllers which possess only local stability guarantees.
%
%
%
%
%
%
}

\textcolor{black}{From a control-theoretic viewpoint, it can be advantageous to have AGAS guarantees for higher-order geometric nonlinear attitude control laws on $\mathrm{SO(3)}$.
%
%
%
This development would have two main benefits.
Firstly, it would extend existing AGAS results for geometric nonlinear PID controllers to more general dynamic compensators on $\mathrm{SO(3)}$.
Secondly, it would enable control designers to use linearization-based design methods in combination with $\mathrm{SO(3)}$-based analysis methods to design and analyze higher-order dynamic compensators for attitude control applications. In this way, practitioners will be able to explicitly take into account various performance requirements and hardware limitations. Moreover, by going beyond PID control, they will be able to exploit additional degrees of freedom in the controller space, thereby designing higher-order controllers, with AGAS guarantees, that perform better than PID controllers at least locally, that is, in the vicinity of the desired equilibrium.}

In recent years, 
there have been some promising developments regarding higher-order geometric nonlinear estimation and control laws.
For the attitude control problem on $\mathrm{SO(3)}$, \cite{invernizzi2022global} proposed a cascaded control architecture with an inner velocity loop and an outer attitude loop.
The attitude loop uses proportional control, whereas the inner loop admits higher-order dynamic compensators. 
Another pertinent result is the development of a higher-order nonlinear complementary filter for kinematic systems on matrix Lie groups \cite{zlotnik2018higher}.
In particular, this filter generalizes the nonlinear complementary filter, proposed in \cite{mahony2008nonlinear}, which is used extensively in attitude estimation applications.
%
%
Motivated by the developments in \cite{invernizzi2022global} and \cite{zlotnik2018higher}, in this paper, we seek analogous results for the attitude control problem, namely, a higher-order geometric nonlinear controller which includes geometric nonlinear PID control as a special case, and which allows for \textcolor{black}{higher-order} dynamic compensation in \emph{both} the attitude and velocity loops.




In particular, this paper presents a theoretical framework for analyzing the stability of higher-order dynamic compensators for attitude control using rotation matrices.
%
%
Preliminary results in this direction have studied the closed-loop stability of a higher-order dynamic state-feedback controller for the rotational double integrator. 
More precisely, in \cite{aslam2022geometrization}, the authors considered a higher-order linear controller for the translational double integrator $J\Ddot{\theta}=u$, \emph{geometrized} it by replacing the angular correction $\theta$ with $\sin(\theta)$, that is, with the gradient of the chordal metric $\Psi(\theta)=1-\cos(\theta)$, and obtained almost global asymptotic stability guarantees for the resulting higher-order geometric nonlinear control law. 
In this paper, our objective is to generalize these results to the case of higher-order geometric nonlinear control on the Special Orthogonal Group $\mathrm{SO(3)}$.

To this end, we present sufficient conditions, based on Lyapunov stability theory, which ensure that the desired equilibrium of the closed-loop tracking error system is rendered AGAS. The candidate Lyapunov function is inspired by quadratic Lyapunov functions of the form $V(x)=x^{\top}Px$ typically considered in the analysis of linear time-invariant (LTI) systems. The analysis proceeds in two steps. In the first step, we obtain a sufficient condition for the positive definiteness of the candidate Lyapunov function, and a necessary and sufficient condition for the negative definiteness of the corresponding Lyapunov rate for almost all initial conditions except those corresponding to the undesired equilibria. In the second step, we relax these conditions to obtain sufficient conditions expressed in the form of linear matrix inequalities (LMIs). Our overall approach is motivated by the widespread use of LMI-based analysis and design tools for LTI systems. To help reduce conservatism, we consider matrix gains for the controller gains as well as the Lyapunov function coefficients. Moreover, we draw parallels, wherever possible, with the necessary and sufficient conditions obtained using a quadratic Lyapunov function for the corresponding linearized system.

We hope that the results presented in this paper will enable control designers to design and analyze higher-order geometric nonlinear controllers for attitude control applications. In particular, by checking the feasibility of the given LMIs, practitioners will be able to confirm that the geometrized version of a given linearization-based dynamic compensator almost globally asymptotically stabilizes the desired equilibrium of the closed-loop attitude control system. The rest of the discussion is structured as follows: after Section \ref{sec:ProblemFormulation} sets up the attitude tracking problem, Section \ref{sec:MainResults} presents the analytical framework for analyzing the stability of a given higher-order geometric nonlinear dynamic compensator. Thereafter, Section \ref{sec:SimulationStudy} demonstrates the applicability of the proposed framework to practical problems such as the design and analysis of robust attitude control laws for multicopters. Finally, Section \ref{sec:Conclusion} concludes the discussion, and outlines some directions for further research. 



\section{Preliminaries}
For vectors $x,y\in\mathbb{R}^n$, $x\cdot y=x^{\top}y$ denotes the inner (or dot) product, and $||x||=\sqrt{x\cdot x}$ denotes the Euclidean norm. For a matrix $M\in\mathbb{R}^{n\times m}$, $||M||$ denotes the induced $2$-norm given by $||M||=\sqrt{\lambda_{\max}(M^{\top}M)}$, where $\lambda_{\max}(A)$ represents the largest eigenvalue of $A\in\mathbb{R}^{n\times n}$. A function $f:\mathbb{R}\rightarrow\mathbb{R}$ belongs to the $\mathcal{L}_p$ space for $p\in[1,\infty)$ if the following $p$-norm of the function exists:
\begin{equation*}
{||f||}_p = {\left\{ \int_0^{\infty} {|f(\tau)|}^p d\tau \right\}}^{1/p}.
\end{equation*}
The Special Orthogonal group $\mathrm{SO(3)}$ is the set of real, $3\times3$, orthogonal
matrices with determinant $1$, that is,
\begin{align*}
\mathrm{SO(3)} & =\left\{ R\in\mathbb{R}^{3\times3}:R^{\top}R=I,\det\left(R\right)=1\right\}.
\end{align*}
For vectors $x,y\in\mathbb{R}^3$, the \emph{cross} product $x\times y=x^{\times}y$, where
\[
x^{\times}={\begin{bmatrix}x_1\\x_2\\x_3\end{bmatrix}}^{\times}=\begin{bmatrix}0&-x_3&x_2\\x_3&0&-x_1\\-x_2&x_1&0\end{bmatrix}.
\]
Using the axis-angle representation and the Rodrigues formula, we can express $R\in\mathrm{SO(3)}$ as:
\begin{equation}
R=\text{exp}(\theta\alpha^{\times})=I+(\sin\theta)\alpha^{\times}+(1-\cos\theta)\alpha^{\times}\alpha^{\times},\label{eq:Re_Axang}
\end{equation}
where $\theta\in\mathbb{R}$ represents the angle of rotation about the axis $\alpha\in\mathbb{S}^{2}$, and $\mathbb{S}^2$ denotes the unit $2$-sphere.
The \emph{vee} $(\vee)$ operator denotes the inverse of the cross $(\times)$ operator, and extracts the entries of the vector $x$ from the skew-symmetric matrix $x^{\times}$, that is, $(x^{\times})^{\vee}=x$. In particular, the cross and vee operators satisfy the following identities \cite{lee2011geometric}:
\begin{subequations}
\begin{align}
x^{\times}y & =x\times y=-y\times x=-y^{\times}x,\label{eq:Identity1}\\
\text{tr}[A x^{\times}] & =-x\cdot(A-A^{\top})^{\vee},\label{eq:Identity3}\\
x^{\times}A+A^{\top}x^{\times} & =\left(\left\{ \text{tr}[A]I-A\right\} x\right)^{\times},\label{eq:Identity4}\\
R x^{\times}R^{\top} & =\left(Rx\right)^{\times},\label{eq:Identity5}
\end{align}
\end{subequations}
for any $x,y\in\mathbb{R}^{3}$, $A\in\mathbb{R}^{3\times3}$, and
$R\in \mathrm{SO(3)}$.

\textcolor{black}{
Configuration error functions play a crucial role in the design of PD controllers for stabilization and tracking on Lie groups (\cite{bullo2019geometric}, Section 11.4). We recall the following definition (\cite{bullo2019geometric}, Definition 11.3).}

\begin{definition}
\textcolor{black}{\emph{(Configuration error function):} A smooth function $\Psi: \mathrm{Q} \rightarrow \mathrm{R}$ is a configuration error function about $q_0 \in \mathrm{Q}$ if it is proper, bounded from below, and if $\Psi(q_0)=0$, $\mathrm{d}\Psi(q_0)=0$, and $\mathrm{Hess}\;\Psi(q_0)$ is positive-definite.}
\end{definition}

\textcolor{black}{As detailed in (\cite{bullo2019geometric}, pg. 531), an immediate consequence of this definition is that $\Psi$ is locally positive-definite about $q_0$. For the Lie group $\mathrm{SO(3)}$, Bullo and Lewis consider the following error function (\cite{bullo2019geometric}, Proposition 11.31):
\begin{align}
\Psi_{\mathrm{SO(3)}}(R) &\isdef \frac{1}{2} \text{tr}[G(I-R)], & G &= \text{tr}[K]I-K,
\end{align}
where $K \in \mathbb{R}^{3\times3}$ is symmetric positive-definite, and $R \in \mathrm{SO(3)}$. As noted in (\cite{bullo2019geometric}, Proposition 11.31), $\Psi_{\mathrm{SO(3)}}$ is a configuration error function about $I$, the identity element of $\mathrm{SO(3)}$. Consequently, in light of (\cite{bullo2019geometric}, Definition 11.3), it is locally positive-definite about $I$.}

\textcolor{black}{In the following discussion, we consider two configuration (or attitude) error functions on $\mathrm{SO(3)}$. The first is the well-known chordal metric:
\begin{equation}
\Psi(R) = {\left. \Psi_{\mathrm{SO(3)}}(R) \right|}_{K=I} = \text{tr}[I-R].\label{eq:ChordalMetric}
\end{equation}
Several important properties of the chordal metric have been succinctly summarized in (\cite{bullo2019geometric}, Proposition 11.31) and (\cite{goodarzi2013geometric}, Proposition 1). These include positive-definiteness, critical points, the left-trivialized derivative, and lower and upper bounds in terms of the squared 2-norm of the left-trivialized derivative.}

\textcolor{black}{In addition to the chordal metric \eqref{eq:ChordalMetric}, we consider a second configuration error function on $\mathrm{SO(3)}$. The latter function was proposed in \cite{lee2011geometric}, \cite{lee2012exponential} to address a potential drawback of the chordal metric, namely, that the attitude error correction obtained using the gradient of the chordal metric approaches zero as the angle of rotation approaches $\pi$ radians. In turn, this behavior can lead to sluggish controller performance, particularly for large-angle rotational errors. The following configuration error function was proposed in \cite{lee2011geometric}, \cite{lee2012exponential} to address this limitation:
\begin{equation}
\Psi_{q}(R) = 2-\sqrt{1+\text{tr}[R]}.\label{eq:AggressiveMetric}
\end{equation}
As pointed out in \cite{lee2012exponential}, this error function is equivalent to a configuration error function that has been used in several quaternion-based attitude control systems. In particular, the quaternion version has been used to develop attitude control laws which ensure that the shortest path to the desired orientation is taken, and the unwinding phenomenon is avoided \cite[Section 7.2]{markley2014fundamentals}.
}

Lastly, the following lemma \cite[Lemma 2.5]{duan2013lmis} plays a key role in the analytical framework developed in this paper.

\begin{lemma}\label{lem:VariableEliminationLemma}
\emph{(Variable Elimination Lemma):} Consider the following restriction set:
\begin{equation*}
\textcolor{black}{\mathcal{F}} \isdef \{F:F\in\mathbb{R}^{n\times n}, F^{\top}F\leq I\}.
\end{equation*}
Let $X\in\mathbb{R}^{m\times n}$, $Y\in\mathbb{R}^{n\times m}$, and $Q\in\mathbb{R}^{m\times m}$. Then:
\begin{equation*}
Q+XFY+Y^{\top}F^{\top}X^{\top}<0,\quad\forall F\in\textcolor{black}{\mathcal{F}} ,
\end{equation*}
if and only if there exists a scalar $\delta>0$, such that
\begin{equation*}
Q+\delta XX^{\top}+\frac{1}{\delta}Y^{\top}Y<0.
\end{equation*}
\end{lemma}


\section{Problem Formulation}\label{sec:ProblemFormulation}
Rigid-body rotational motion is governed by the following equations of motion \textcolor{black}{(\cite{marsden1999introduction}, Chapter 15)}:
\begin{equation}
\begin{split}
\dot{R} & =R\omega^{\times},\\
J\dot{\omega} & =-\omega^{\times}J\omega+\tau.
\end{split}
\label{eq:EoMs}
\end{equation}
Here, $R\in \mathrm{SO(3)}$ denotes the attitude (or orientation) of the body-fixed reference frame relative to the inertial reference frame, $\omega\in\Rset^3$ denotes the body angular velocity in body coordinates, $J\in\Rset^{3\times3}$ denotes the symmetric positive definite inertia matrix, and $\tau\in\Rset^3$ denotes the control torque. The control problem is to track a desired (or reference) attitude $R_d\in \mathrm{SO(3)}$ with the following kinematics:
\begin{equation}
\dot{R}_{d}=R_{d}\omega_{d}^{\times},\label{eq:Rd_dot}
\end{equation}
where $\omega_{d}$ denotes the desired angular velocity. We assume that $\omega_d$ is bounded and continuously differentiable, with bounded derivative $\dot{\omega}_d$.

In order to formulate the attitude tracking problem, we consider the following attitude and velocity tracking errors:
\begin{align}
R_{e} & \isdef R_{d}^{\top}R, & \omega_{e}\isdef\omega-R_{e}^{\top}\omega_{d}.\label{eq:Errors}
\end{align}
Using these, we obtain the tracking error system as:
\begin{equation}
\begin{split}\dot{R}_{e} & =R_{e}\omega_{e}^{\times},\\
J\dot{\omega}_{e} & =-\omega^{\times}J\omega+\tau-J\dot{\omega}_v,\quad\quad\omega_v\isdef R_e^{\top}\omega_d.
\end{split}
\label{eq:ErrorSystem_A}
\end{equation}
For the attitude error function, we choose the chordal (or Euclidean) metric:
\begin{equation}
\Psi(R_{e})\isdef\frac{1}{2}\text{tr}[I-R_{e}].\label{eq:Psi}
\end{equation}
The corresponding attitude error vector is given by 
\begin{equation}
e_{R}\isdef\frac{1}{2}(R_{e}-R_{e}^{\top})^{\vee}.\label{eq:eR}
\end{equation}

Next, we consider a control law that \textcolor{black}{applies a torque consisting of} a cancellation term and a feedback stabilization term, where the latter is obtained using a higher-order dynamic compensator. More precisely, the torque applied is
\begin{equation}
\tau=u_{C}+u,\label{eq:tau}
\end{equation}
where $u_{C}$ denotes the cancellation torque
\begin{align}
u_{C} & \isdef\omega^{\times}J\omega+J\dot{\omega}_{v}, & \omega_{v}=R_{e}^{\top}\omega_{d},\label{eq:uC}
\end{align}
and $u$ is the output generated by a dynamic compensator, \textcolor{black}{with state $x_K \in \mathbb{R}^n$}, obtained as:
\begin{equation}
\begin{split}\dot{x}_{K} & =A_{K}x_{K}+B_{\theta}e_{R}+B_{\omega}\omega_{e},\\
u & =C_{K}x_{K}+D_{\theta}e_{R}+D_{\omega}\omega_{e}.
\end{split}
\label{eq:Controller}
\end{equation}
We assume that the controller matrices $A_{K}\in\Rset^{n\times n}$, $(B_{\theta},B_{\omega})\in\Rset^{n\times3}$, $C_{K}\in\Rset^{3\times n}$, and $(D_{\theta},D_{\omega})\in\Rset^{3\times3}$ are known, that is, we assume that a controller of the form \eqref{eq:Controller} has already been designed. We also assume that the state space realization,
\begin{equation}
\Sigma_K\isdef\left[\begin{array}{c|c|c}
             A_K & B_{\theta} & B_{\omega} \\ \hline
             C_K & D_{\theta} & D_{\omega}
           \end{array}\right],
\end{equation}
constitutes a minimal realization.
Substituting the control law \eqref{eq:tau}-\eqref{eq:Controller} in \eqref{eq:ErrorSystem_A}, we obtain the following closed-loop tracking error system:
\begin{equation}
\begin{split}\dot{R}_{e} & =R_{e}\omega_{e}^{\times},\\
J\dot{\omega}_{e} & =C_{K}x_{K}+D_{\theta}e_{R}+D_{\omega}\omega_{e},\\
\dot{x}_{K} & =A_{K}x_{K}+B_{\theta}e_{R}+B_{\omega}\omega_{e}.
\end{split}
\label{eq:ClosedLoop}
\end{equation}

\begin{problem}\label{prob:MainProblem}
\textcolor{black}{Obtain sufficient conditions, expressed in the form of linear matrix inequalities (LMIs), which ensure that the desired equilibrium $(R_{e},\omega_{e},x_{K})=(I,0,0)$ of the closed-loop tracking error system \eqref{eq:ClosedLoop} is almost globally asymptotically stable (AGAS).}
\end{problem}


Before presenting our solution to Problem \ref{prob:MainProblem}, let us first consider the linearized tracking error system corresponding to \eqref{eq:ClosedLoop}:
\begin{equation}
\begin{split}\dot{\xi} & =\eta,\\
J\dot{\eta} & =C_{K}x_{K}+D_{\theta}\xi+D_{\omega}\eta,\\
\dot{x}_K & =A_{K}x_{K}+B_{\theta}\xi+B_{\omega}\eta.
\end{split}
\label{eq:ClosedLoop_Linearized}
\end{equation}
In particular, the linearized system uses the small angle assumption, $R_e\approx I+\xi^{\times}$, and defines the approximate velocity error as $\eta=\omega-\omega_d$. We analyze its stability using the following candidate Lyapunov function:
\begin{equation}
V_{L}\isdef x^{\top}\mathcal{P}_{L}x,\label{eq:V_lin}
\end{equation}
where
\begin{align}
x &=\begin{bmatrix}\xi\\\eta\\x_{K}\end{bmatrix}, & \mathcal{P}_{L}=\begin{bmatrix}P_{11} & * & *\\
JP_{21} & P_{22}J & *\\
P_{31} & P_{32}J & P_{33}
\end{bmatrix},\label{eq:P_lin}
\end{align}
and where $*$ denotes the transposed version of the corresponding off-diagonal term. Suppose that the Lyapunov coefficient matrix $\mathcal{P}_{L}$ and the Lyapunov rate $\dot{V}_{L}$ along trajectories of \eqref{eq:ClosedLoop_Linearized} satisfy the following conditions:
\begin{align*}
\mathcal{P}_{L}=\mathcal{P}_{L}^{\top} & >0, & \dot{V}_{L} & <0.
\end{align*}
Then, we can conclude that the controller \eqref{eq:Controller} locally asymptotically stabilizes the desired equilibrium $(R_{e},\omega_{e},x_{K})=(I,0,0)$ of the tracking error system \eqref{eq:ClosedLoop}. In the next section, we will consider a candidate Lyapunov function $V$ whose coefficients are a special case of the coefficients used in \eqref{eq:P_lin}. Moreover, for this Lyapunov function, we will obtain: $\left.1\right)$ a sufficient condition which ensures that $V>0$, and $\left.2\right)$ a necessary and sufficient condition which ensures that $\dot{V}<0$. We will then use an upper bound for the Lyapunov rate to obtain sufficient conditions, in the form of linear matrix inequalities (LMIs), which ensure that the desired equilibrium is rendered AGAS by a given dynamic compensator of the form \eqref{eq:Controller}.

\begin{remark}
\textcolor{black}{The dynamic compensator \eqref{eq:Controller} is a general framework that models several existing geometric nonlinear PID designs as special cases.
These include the PID controllers proposed in \cite{goodarzi2013geometric}, \cite{aslam2023CDC}, as well as the cascade P/PI and P/PID architectures analyzed in \cite{invernizzi2020robust}, \cite{invernizzi2022global}. 
Section \ref{sec:PID} discusses these PID variants in more detail. 
}
\end{remark}

\begin{remark}
\textcolor{black}{
The sufficient conditions for several PD and PID controllers proposed in the $\mathrm{SO(3)}$ literature, such as those in \cite{lee2010geometric}, \cite{goodarzi2013geometric}, can equivalently be expressed as linear matrix inequalities (LMIs). The use of LMIs over a non-Euclidean space such as $\mathrm{SO(3)}$ is possible if the matrices used in the quadratic forms which bound the Lyapunov function candidate and its time derivative 
are linear functions of the Lyapunov coefficients. In addition, feasibility problems involving LMIs can often be solved using convex optimization \cite{boyd1994linear}.
}
\end{remark}

\begin{remark}\label{Remark:LinearDamping}
The problem formulation can be extended to include a linear damping term with known damping matrix. In particular, consider the following modified attitude dynamics:
\[
J\dot{\omega}=-\omega^{\times}J\omega-\Gamma\omega+\tau,
\]
where $\Gamma=\Gamma^{\top}\in\Rset^{3\times3}$ denotes a positive semidefinite damping matrix. Consider also the following modified cancellation torque:
\begin{align}
u_{C} & =\omega^{\times}J\omega+J\dot{\omega}_{v}+\Gamma R_e^{\top}\omega_d, & \omega_{v}=R_{e}^{\top}\omega_{d}.\label{eq:uC_Modified}
\end{align}
Then, the closed-loop tracking error system, with dynamic compensation \eqref{eq:Controller}, is given by:
\begin{equation}
\begin{split}\dot{R}_{e} & =R_{e}\omega_{e}^{\times},\\
J\dot{\omega}_{e} & =C_{K}x_{K}+D_{\theta}e_{R}+(D_{\omega}-\Gamma)\omega_{e},\\
\dot{x}_{K} & =A_{K}x_{K}+B_{\theta}e_{R}+B_{\omega}\omega_{e}.
\end{split}\label{eq:ClosedLoop_Modified}
\end{equation}
Comparing \eqref{eq:ClosedLoop} and \eqref{eq:ClosedLoop_Modified}, we note that the damping matrix for the closed-loop tracking-error system has changed from $D_{\omega}$ to $D_{\omega}-\Gamma$. In the following discussion, we will present stabilization conditions for the error system \eqref{eq:ClosedLoop}, whilst noting that systems with known linear damping can easily be included in the proposed analytical framework.
\end{remark}

\section{Main Results}\label{sec:MainResults}
\subsection{Attitude Error Function}

The time derivative of the attitude error function \eqref{eq:Psi} is given by:
\[
\dot{\Psi} =-\frac{1}{2}\text{tr}[\dot{R}_e]=-\frac{1}{2}\text{tr}[R_e\omega_e^{\times}]=\frac{1}{2}(R_e-R_e^{\top})^{\vee}\cdot\omega_e.
\]
In the above simplification, we have used the error kinematics \eqref{eq:ClosedLoop} as well as the identity \eqref{eq:Identity3}. 
Furthermore, using the attitude error vector \eqref{eq:eR}, we observe that:
\begin{equation}
\dot{\Psi} =e_{R}\cdot\omega_e.\label{eq:Psidot}
\end{equation}
Similarly, we can obtain the rate of change of the attitude error vector as:
\begin{align}
\dot{e}_{R}^{\times} & =\frac{1}{2}(\dot{R}_e-\dot{R}_e^{\top})=\frac{1}{2}(R_e\omega_e^{\times}+\omega_e^{\times}R_e^{\top})\nonumber \\
 & =\frac{1}{2}\{(\text{tr}[R_e]I-R_e^{\top})\omega_e\}^{\times}\nonumber \\
\implies\dot{e}_{R} & =E(R_e)\omega_e,\label{eq:eR_dot}
\end{align}
where the mapping $E:\mathbb{R}^{3\times3}\rightarrow\mathbb{R}^{3\times3}$ is given as follows:
\begin{equation}
E(M)\isdef\frac{1}{2}(\text{tr}[M]I-M^{\top}),\label{eq:E}
\end{equation}
and we have used the identity \eqref{eq:Identity4}.

\subsection{Lyapunov Stability Analysis}\label{sec:LyapunovAnalysis_Chordal}

Using the attitude error function \eqref{eq:Psi}, \textcolor{black}{we define the following Lyapunov function candidate over the space $\mathrm{SO(3)} \times \mathbb{R}^3 \times \mathbb{R}^n$}:
\begin{equation}
\begin{split}V(R_e,\omega_e,x_K)\isdef & \;2p_{11}\Psi(R_e)+\omega_{e}\cdot P_{22}J\omega_{e}\\
 & \;+2e_{R}\cdot P_{21}^{\top}J\omega_{e}+x_{K}\cdot P_{33}x_{K}\\
 & \;+2x_{K}\cdot P_{31}e_{R}+2x_{K}\cdot P_{32}J\omega_{e},
\end{split}
\label{eq:V}
\end{equation}
where $p_{11}\in\mathbb{R}$, 
$(P_{21},P_{22})\in\mathbb{R}^{3\times3}$, $(P_{31},P_{32})\in\mathbb{R}^{n\times3}$, and 
$P_{33}\in\mathbb{R}^{n\times n}$. 
From \cite[Proposition 1]{goodarzi2013geometric}, we have the following lower bound for the attitude error function:
\[
\Psi\geq\frac{1}{2}||e_{R}||^{2}.
\]
Using this relation, we bound $V$ as:
\[
\begin{split}V\geq & \;p_{11}||e_R||^2+\omega_{e}\cdot P_{22}J\omega_{e}+2e_{R}\cdot P_{21}^{\top}J\omega_{e}\\
 & \quad\;+x_{K}\cdot P_{33}x_{K}+2x_{K}\cdot P_{31}e_{R}+2x_{K}\cdot P_{32}J\omega_{e}.
\end{split}
\]
Consequently, a sufficient condition for $V$ to be positive definite can be stated as:
\begin{equation}
\mathcal{P}\isdef\begin{bmatrix}p_{11}I & * & *\\
JP_{21} & P_{22}J & *\\
P_{31} & P_{32}J & P_{33}
\end{bmatrix}>0.\label{eq:V_PosDef}
\end{equation}
\textcolor{black}{Comparing $\mathcal{P}$ with $\mathcal{P}_L$ in \eqref{eq:P_lin}, we note that the only difference is that the matrix coefficient $P_{11}\in\mathbb{R}^{3\times3}$ in \eqref{eq:P_lin} has been replaced by $p_{11}I$.}

\begin{remark}
\textcolor{black}{The Lyapunov function candidate \eqref{eq:V} includes, as special cases, the Lyapunov functions considered in earlier work \cite{lee2010geometric,goodarzi2013geometric} on the design of PD and PID controllers on $\mathrm{SO(3)}$. For instance, the Lyapunov function used in \cite{lee2010geometric} to develop a geometric nonlinear PD controller on $\mathrm{SO(3)}$ can be expressed as:
\begin{equation}
\begin{split}V(R_e,\omega_e)\isdef & \;k_R\Psi(R_e)+\frac{1}{2}\omega_{e}\cdot J\omega_{e}+c_2 e_{R} \cdot \omega_{e},
\end{split}\label{eq:V_Lee}
\end{equation}
where $(c_2,k_R)$ are positive scalars. We observe that \eqref{eq:V_Lee} is a special case of \eqref{eq:V}, and can be obtained by choosing the Lyapunov coefficients as $p_{11} = k_R/2$, $P_{22} = (1/2)I$, $P_{21} = (c_2/2)J^{-1}$, $P_{33} = 0$, and $P_{31} = P_{32} = 0$. 
}
\end{remark}

\begin{remark}
\textcolor{black}{The Lyapunov function candidate \eqref{eq:V} is closely related to the energy functions considered by Koditschek \cite{koditschek1989application} and by Bullo and Lewis (\cite{bullo2019geometric}, Lemma 11.33) for designing PD controllers on $\mathrm{SO(3)}$. In particular, the aforementioned PD controllers are designed using Lyapunov functions which modify the total energy of the system. The total energy is given by the sum of a configuration (or attitude) error function and the rotational kinetic energy, and the modification to the total energy function consists, essentially, of the addition of a term involving the dot product of the attitude error vector $e_R$ and the angular velocity error $\omega_e$. This idea is sometimes referred to as``Chetaev's trick" (\cite{bullo2019geometric}, pg. 343). The Lyapunov function candidate \eqref{eq:V} extends this idea by including a positive-definite term for the controller state $x_K$, and corresponding cross-terms for the pairs $(x_K,e_R)$ and $(x_K,\omega_e)$.
}
\end{remark}

\begin{proposition}\label{prop:Vdot}
Consider the closed-loop tracking error system \eqref{eq:ClosedLoop} and the candidate Lyapunov function \eqref{eq:V}. The Lyapunov rate along trajectories of \eqref{eq:ClosedLoop} is given by:
\begin{equation}
\begin{split}
\dot{V} & =x^{\top}\mathcal{M}_0x+2\omega_{e}\cdot JP_{21}E(R_{e})\omega_{e}\\
 & \quad\;+2x_{K}\cdot P_{31}E(R_{e})\omega_{e},
\end{split}\label{eq:Vdot}
\end{equation}
where
\begin{align}
x & \isdef\begin{bmatrix}e_{R}\\
\omega_{e}\\
x_{K}
\end{bmatrix}, & \mathcal{M}_0\isdef\begin{bmatrix}M_{11} & * & *\\
M_{21} & M_{22} & *\\
M_{31} & M_{32} & M_{33}
\end{bmatrix},\label{eq:M0}
\end{align}
the matrix $E(R_e)$ is obtained using \eqref{eq:E}, and the submatrices $M_{ij}$ are defined as follows:
\begin{equation}
\begin{split}
M_{11} & \isdef(P_{21}^{\top}D_{\theta}+D_{\theta}^{\top}P_{21})+(P_{31}^{\top}B_{\theta}+B_{\theta}^{\top}P_{31}),\\
M_{22} & \isdef(P_{22}D_{\omega}+D_{\omega}^{\top}P_{22})+(JP_{32}^{\top}B_{\omega}+B_{\omega}^{\top}P_{32}J),\\
M_{21} & \isdef p_{11}I+P_{22}D_{\theta}+D_{\omega}^{\top}P_{21}+JP_{32}^{\top}B_{\theta}+B_{\omega}^{\top}P_{31},\\
M_{33} & \isdef (P_{32}C_{K}+C_{K}^{\top}P_{32}^{\top})+(P_{33}A_{K}+A_{K}^{\top}P_{33}),\\
M_{31} & \isdef P_{32}D_{\theta}+C_{K}^{\top}P_{21}+A_{K}^{\top}P_{31}+P_{33}B_{\theta},\\
M_{32} & \isdef P_{32}D_{\omega}+C_{K}^{\top}P_{22}+A_{K}^{\top}P_{32}J+P_{33}B_{\omega}.
\end{split}\label{eq:Mij}
\end{equation}
\end{proposition}

\begin{proof}
The time derivative of $V$ is given by:
\begin{align*}
\dot{V} & =2p_{11}\dot{\Psi}+2\omega_{e}\cdot P_{22}J\dot{\omega}_{e}+2e_{R}\cdot P_{21}^{\top}J\dot{\omega}_{e}\\
 & \quad\;+2\dot{e}_{R}\cdot P_{21}^{\top}J\omega_{e}+2x_{K}\cdot P_{33}\dot{x}_{K}+2x_{K}\cdot P_{31}\dot{e}_{R}\\
 & \quad\;+2\dot{x}_{K}\cdot P_{31}e_{R}+2x_{K}\cdot P_{32}J\dot{\omega}_{e}+2\dot{x}_{K}\cdot P_{32}J\omega_{e}\\
 & =2p_{11}\dot{\Psi}+2(P_{21}e_{R}+P_{22}\omega_{e}+P_{32}^{\top}x_{K})\cdot J\dot{\omega}_{e}\\
 & \quad\;+2(P_{31}e_{R}+P_{32}J\omega_{e}+P_{33}x_{K})\cdot\dot{x}_{K}\\
 & \quad\;+2\dot{e}_{R}\cdot P_{21}^{\top}J\omega_{e}+2x_{K}\cdot P_{31}\dot{e}_{R}.
\end{align*}
Substituting \eqref{eq:ClosedLoop}, \eqref{eq:Psidot}, and \eqref{eq:eR_dot}, we express the Lyapunov rate as:
\begin{align*}
\dot{V} & =2p_{11}e_{R}\cdot\omega_{e}+2(P_{21}e_{R})\cdot(C_{K}x_{K}+D_{\theta}e_{R}+D_{\omega}\omega_e)\\
 & \quad\;+2(P_{22}\omega_{e}+P_{32}^{\top}x_{K})\cdot(C_{K}x_{K}+D_{\theta}e_{R}+D_{\omega}\omega_e)\\
 & \quad\;+2(P_{31}e_{R}+P_{32}J\omega_{e})\cdot(A_{K}x_{K}+B_{\theta}e_{R}+B_{\omega}\omega_e)\\
 & \quad\;+2x_{K}\cdot P_{33}(A_{K}x_{K}+B_{\theta}e_{R}+B_{\omega}\omega_e)\\
 & \quad\;+2\omega_{e}\cdot JP_{21}E(R_{e})\omega_{e}+2x_{K}\cdot P_{31}E(R_{e})\omega_{e}.
\end{align*}
Re-arranging terms, we have that:
\begin{align*}
\dot{V} & =2e_{R}\cdot P_{21}^{\top}D_{\theta}e_{R}+2e_{R}\cdot P_{31}^{\top}B_{\theta}e_{R}+2\omega_{e}\cdot P_{22}D_{\omega}\omega_e\\
 & \quad\;+2\omega_{e}\cdot JP_{32}^{\top}B_{\omega}\omega_e+2p_{11}e_{R}\cdot\omega_{e}+2\omega_{e}\cdot P_{22}D_{\theta}e_{R}\\
 & \quad\;+2e_{R}\cdot P_{21}^{\top}D_{\omega}\omega_e+2\omega_{e}\cdot JP_{32}^{\top}B_{\theta}e_{R}\\
 & \quad\;+2e_{R}\cdot P_{31}^{\top}B_{\omega}\omega_e+2x_{K}\cdot P_{32}C_{K}x_{K}\\
 & \quad\;+2x_{K}\cdot P_{33}A_{K}x_{K}+2x_{K}\cdot P_{32}D_{\theta}e_{R}\\
 & \quad\;+2e_{R}\cdot P_{21}^{\top}C_{K}x_{K}+2e_{R}\cdot P_{31}^{\top}A_{K}x_{K}\\
 & \quad\;+2x_{K}\cdot P_{33}B_{\theta}e_{R}+2x_{K}\cdot P_{32}D_{\omega}\omega_e\\
 & \quad\;+2\omega_{e}\cdot P_{22}C_{K}x_{K}+2\omega_{e}\cdot JP_{32}^{\top}A_{K}x_{K}\\
 & \quad\;+2x_{K}\cdot P_{33}B_{\omega}\omega_e+2\omega_{e}\cdot JP_{21}E(R_{e})\omega_{e}\\
 & \quad\;+2x_{K}\cdot P_{31}E(R_{e})\omega_{e}.
\end{align*}
Lastly, the Lyapunov rate \eqref{eq:Vdot} is obtained by substituting the submatrices $M_{ij}$ defined in \eqref{eq:Mij}. 
\end{proof}

\begin{proposition}
The Lyapunov rate \eqref{eq:Vdot} is negative definite if and only if there exist positive scalars $\tau_1$ and $\tau_2$ such that the following matrix inequality is satisfied:
\begin{equation}
\mathcal{M}_1\isdef\begin{bmatrix}M_{11} & * & *\\
M_{21} & \tilde{M}_{22} & *\\
M_{31} & M_{32} & \tilde{M}_{33}
\end{bmatrix}<0,\label{eq:Vdot_NegDef}
\end{equation}
where
\begin{equation}
\begin{split}
\tilde{M}_{22} &\isdef M_{22}+\tau_1I+\tau_2I+\frac{1}{\tau_2}J(P_{21}P_{21}^{\top})J,\\
\tilde{M}_{33} &\isdef M_{33}+\frac{1}{\tau_1}(P_{31}P_{31}^{\top}).
\end{split}\label{eq:Mtilde}
\end{equation}
\end{proposition}

\begin{proof}
The last two terms in \eqref{eq:Vdot} contain the contribution of the time derivative \eqref{eq:eR_dot} of the attitude error vector, that is, $\dot{e}_R=E(R_e)\omega_e$. Using the axis-angle representation \eqref{eq:Re_Axang} of the attitude error $R_e$, we express the eigenvalues of the matrix $E^{\top}(R_e)E(R_e)$ as
\[
\frac{1}{2}(1+\cos\theta_e),\quad\frac{1}{2}(1+\cos\theta_e),\quad\cos^2\theta_e.
\]
Since $E^{\top}(R_e)E(R_e)\leq I$, we can apply the variable elimination lemma (Lemma
\ref{lem:VariableEliminationLemma}) 
to the last two terms in \eqref{eq:Vdot}. In particular, applying this lemma to the last term in \eqref{eq:Vdot}, we note that the requirement $\dot{V}<0$ is satisfied if and only if there exists a scalar $\tau_1>0$ such that
\begin{equation}
\begin{split}
& x^{\top}\mathcal{M}_0x+\frac{1}{\tau_1}x_{K}^{\top}(P_{31}P_{31}^{\top})x_{K}+\tau_1\omega_{e}^{\top}\omega_{e}\\
 & \quad\;+2\omega_{e}\cdot JP_{21}E(R_{e})\omega_{e}<0.
\end{split}\label{eq:Vdot_B}
\end{equation}
Next, we apply the variable elimination lemma to the remaining term containing the matrix $E(R_e)$, and note that the condition stipulated in \eqref{eq:Vdot_B} holds if and only if there exists a scalar $\tau_2>0$ such that
\begin{equation}
\begin{split}
& x^{\top}\mathcal{M}_0x+\frac{1}{\tau_1}x_{K}^{\top}(P_{31}P_{31}^{\top})x_{K}+\tau_1\omega_{e}^{\top}\omega_{e}\\
 & \quad\;+\frac{1}{\tau_2}\omega_{e}^{\top}J(P_{21}P_{21}^{\top})J\omega_{e}+\tau_2\omega_{e}^{\top}\omega_{e}<0.
\end{split}\label{eq:Vdot_C}
\end{equation}
Defining the matrices $\tilde{M}_{22}\in\mathbb{R}^{3\times3}$ and $\tilde{M}_{33}\in\mathbb{R}^{n\times n}$ as in \eqref{eq:Mtilde}, we obtain the matrix inequality \eqref{eq:Vdot_NegDef} which constitutes a necessary and sufficient condition for the negative definiteness of the Lyapunov rate \eqref{eq:Vdot}. 
\end{proof}


\begin{theorem}
Consider the closed-loop tracking error system \eqref{eq:ClosedLoop} and the candidate Lyapunov function \eqref{eq:V}. Suppose that the Lyapunov function coefficients satisfy the matrix inequalities \eqref{eq:V_PosDef} and \eqref{eq:Vdot_NegDef}, with the relevant submatrices defined in \eqref{eq:Mij} and \eqref{eq:Mtilde}. Then, \textcolor{black}{the following properties hold:}
\textcolor{black}{
\begin{enumerate}
\item The zero equilibrium of the tracking errors $(e_R,\omega_e,x_K)$ is stable.
\item The tracking errors $(e_R,\omega_e,x_K)$ converge asymptotically to zero, that is, $R_e \rightarrow \{I\} \cup \{\emph{exp}(\pi\alpha^{\times})|\alpha\in\mathbb{S}^2\}$, and $(\omega_e,x_K) \rightarrow 0$ as $t \rightarrow \infty$.
\item The undesired equilibria where $R_e \in \{\emph{exp}(\pi\alpha^{\times})|\alpha\in\mathbb{S}^2\}$ are unstable.
\item The desired equilibrium is almost globally asymptotically stable (AGAS).
\end{enumerate}}
\end{theorem}

\begin{proof}
The condition \eqref{eq:V_PosDef} ensures that the candidate Lyapunov function \eqref{eq:V} is positive definite, whereas \eqref{eq:Vdot_NegDef} ensures that the corresponding Lyapunov rate $\dot{V}$ is negative definite.
\textcolor{black}{Thus, there exist positive definite matrices $(W_1,W_2) \in \mathbb{R}^{3\times3}$ such that $V \geq \xi^{\top}W_1\xi$ and $\dot{V} \leq \xi^{\top}W_2\xi$, where $\xi = [||e_R||;||\omega_e||;||x_K||] \in \mathbb{R}^3$.
As a result, the Lyapunov function $V(t)$ is bounded from below and is nonincreasing. This shows $\emph{(1)}$.}

\textcolor{black}{
In addition, the Lyapunov function has a limit, $\lim_{t \rightarrow \infty} V(t) = V_{\infty}$, and the tracking errors $(e_R,\omega_e,x_K)$ $\in \mathcal{L}_{\infty}$. 
The boundedness of the tracking errors and \eqref{eq:ClosedLoop} ensure the boundedness of the tracking error rates, that is, $(\dot{e}_R,\dot{\omega}_e,\dot{x}_K) \in \mathcal{L}_{\infty}$.
Moreover, since $\int_{0}^{\infty} \xi(\tau)^{\top} W_2$ $\xi(\tau) d\tau \leq V(0) - V_{\infty}$, it follows that the tracking errors $(e_R,\omega_e,x_K) \in \mathcal{L}_2$ as well.
Applying Barbalat's lemma \textcolor{black}{[Khalil]}, we note that $(e_R,\omega_e,x_K) \rightarrow 0$ as $t \rightarrow \infty$.
Let $(\alpha,\theta_e)\in\mathbb{S}^2\times\mathbb{R}$ denote the axis-angle representation \eqref{eq:Re_Axang} corresponding to the attitude error $R_e=\text{exp}(\theta_e\alpha^{\times})$. Then, $e_R=0.5(R_e-R_e^{\top})^{\vee}=(\sin\theta_e)\alpha$. Since $\alpha$ is a unit vector, the asymptotic convergence of $e_R$ to zero implies that 
the attitude error $R_e$ converges either to the desired equilibrium or to the undesired equilibria, that is, $R_e \rightarrow \{I\} \cup \{\text{exp}(\pi\alpha^{\times})|\alpha\in\mathbb{S}^2\}$ as $t \rightarrow \infty$.
This shows $\emph{(2)}$.}
%
%

Next, we use similar arguments as in (\cite{goodarzi2013geometric_arxiv}, Appendix A) \textcolor{black}{and (\cite{lee2013robust}, Proposition 3)} to show that the undesired equilibria $(R_e,\omega_e,x_K)=(\text{exp}(\pi\alpha^\times),0,0)$ are unstable.
To this end, we note that at the undesired equilibria, the attitude error function $\Psi=2$, and the candidate Lyapunov function $V=4p_{11}$. Defining
\begin{equation*}
W\isdef4p_{11}-V,
\end{equation*}
it follows that $W=0$ at the undesired equilibria, and
\begin{equation*}
\begin{split}
W\geq&\;2p_{11}(2-\Psi)-||P_{22}J||||\omega_e||^2-||P_{33}||||x_K||^2\\
&-2||P_{21}J||||e_R||||\omega_e||-2||P_{31}||||x_K||||e_R||\\
&-2||P_{31}J||||x_K||||\omega_e||.
\end{split}
\end{equation*}
Since $\Psi$ is continuous, we can choose the initial attitude $R$ such that in an arbitrary small neighborhood of an undesired equilibrium, it holds that $2-\Psi>0$. Moreover, if $||\omega_e||$ and $||x_K||$ are sufficiently small, then $W>0$ in this neighborhood. Consequently, the existence of Lyapunov function coefficients which ensure that $(V>0,\dot{V}<0)$ also ensures that in any arbitrary small neighborhood of an undesired equilibrium, there exists a domain in which $W>0$, and $\dot{W}=-\dot{V}>0$. Applying Chetaev’s instability theorem \cite[Theorem 4.3]{Khalil2002}, we conclude that the undesired equilibria are unstable. This shows $\emph{(3)}$.
\textcolor{black}{Lastly, we note that the region of attraction to the undesired equilibria has 
zero measure \cite{goodarzi2013geometric_arxiv}, \cite{lee2013robust}.}
Consequently, for almost all initial conditions, the trajectories of the closed-loop error system \eqref{eq:ClosedLoop} converge to the desired equilibrium $(R_{e},\omega_e,x_K)=(I,0,0)$.
%
%
This shows $\emph{(4)}$. $\blacksquare$
\end{proof}

Given a dynamic compensator of the form \eqref{eq:Controller}, the matrix inequalities \eqref{eq:V_PosDef} and \eqref{eq:Vdot_NegDef} can be used to search for Lyapunov function coefficients which yield AGAS guarantees for the closed-loop tracking error system \eqref{eq:ClosedLoop}. We note that all but two of the submatrices in \eqref{eq:V_PosDef} and \eqref{eq:Vdot_NegDef} are linear functions of the Lyapunov function coefficients. \textcolor{black}{However, by considering an upper bound for \eqref{eq:Vdot_C}} 
and using the Schur complement, we can obtain a sufficient condition for the negative definiteness of the Lyapunov rate in which the various submatrices are linear functions of the unknown coefficients, and which can thus be expressed in the form of linear matrix inequalities (LMIs). More precisely, recalling the condition \eqref{eq:Vdot_C} from the proof of Proposition 2, we note that the following constitutes a sufficient condition for the Lyapunov rate to be negative definite:
\begin{equation}
\begin{split}
& x^{\top}\mathcal{M}_0x+x_{K}^{\top}N_3x_{K}+\omega_{e}^{\top}(\tau_1I+\tau_2I+N_2)\omega_{e}<0,\\
&\quad\frac{1}{\tau_2}J(P_{21}P_{21}^{\top})J\leq N_2,\quad\frac{1}{\tau_1}(P_{31}P_{31}^{\top})\leq N_3.
\end{split}\label{eq:Vdot_D}
\end{equation}
Using the Schur Complement Lemma \cite[Lemma 2.6]{duan2013lmis}, we reformulate \eqref{eq:Vdot_D} as:
\begin{equation}
\begin{split}
& \mathcal{M}_2\isdef\begin{bmatrix}M_{11} & * & *\\
M_{21} & M_{22}+(\tau_1+\tau_2)I + N_2 & *\\
M_{31} & M_{32} & M_{33}+N_3
\end{bmatrix} <0,\\
&\quad\quad\quad\begin{bmatrix}N_2 & JP_{21}\\P_{21}^{\top}J & \tau_2 I\end{bmatrix}\geq0,\quad\quad\quad\begin{bmatrix}N_3 & P_{31}\\P_{31}^{\top} & \tau_1 I\end{bmatrix}\geq0.
\end{split}\label{eq:Vdot_NegDef_LMI}
\end{equation}
\begin{corollary}\label{Theorem:Chordal_LMI}
Consider the closed-loop tracking error system \eqref{eq:ClosedLoop} and the candidate Lyapunov function \eqref{eq:V}. Suppose that the Lyapunov function coefficients satisfy the linear matrix inequalities \eqref{eq:V_PosDef} and \eqref{eq:Vdot_NegDef_LMI}, with the relevant submatrices defined in \eqref{eq:Mij}. Then, $V>0$, $\dot{V}<0$, and the controller \eqref{eq:Controller} almost globally asymptotically stabilizes the desired equilibrium $(R_e,\omega_e,x_K)=(I,0,0)$ of the error system \eqref{eq:ClosedLoop}.
\end{corollary}

\begin{remark}
\textcolor{black}{Theorem 1 and Corollary 1 present an analytical framework which can be used to certify the almost global asymptotic stability of a higher-order geometric nonlinear dynamic compensator for attitude control on $\mathrm{SO(3)}$. Such compensators could potentially be designed using linearization-based methods, and could then be analyzed for almost global asymptotic stability using the LMIs proposed in Corollary 1. In this way, practitioners could use frequency domain loopshaping techniques for the linearized system \eqref{eq:ClosedLoop_Linearized} to design attitude controllers which have better (local) performance guarantees than PID controllers. They could then geometrize the designed compensator and use the proposed LMIs to obtain a certificate of almost global asymptotic stability for the resulting higher-order geometric nonlinear controller. Moreover, the class of higher-order dynamic compensators under consideration includes geometric nonlinear PD and PID controllers as special cases. In this way, the proposed framework generalizes existing results, and gives designers more freedom in developing attitude control laws on $\mathrm{SO(3)}$.}
\end{remark}

\begin{remark}
\textcolor{black}{Theorem 1 considers a particular instance of a holonomic mechanical system on a matrix Lie group, namely, attitude control on $\mathrm{SO(3)}$ for a fully-actuated rigid body. In principle, it should be possible to obtain analogous results for more general holonomic mechanical systems on Lie groups, in particular, for the class of systems considered in \cite{maithripala2015intrinsic}. Moreover, for the Lie group $\mathrm{SO(3)}$, the framework can be applied to attitude error functions other than the chordal metric. These include: the attitude error function proposed in \cite{lee2011geometric}, \cite{lee2012exponential}; the logarithmic map (or geodesic metric) on $\mathrm{SO(3)}$; and the attitude error function considered in (\cite{zlotnik2016exponential}, Proposition 1). Lastly, the proposed LMI-based framework in Corollary 1 is, at the moment, only a verification tool and not a design tool. More precisely, once a higher-order controller has been designed, its almost global asymptotic stability can be verified using the LMIs in Corollary 1. In future work, we hope to add design considerations to the problem formulation, and to obtain a LMI-based synthesis method for attitude control on $\mathrm{SO(3)}$.
}
\end{remark}

\subsection{Changing the Attitude Error Function}\label{subsec:AggressiveMetric}

Up to this point, the analytical framework has been developed using the chordal metric \eqref{eq:Psi}, the attitude error vector \eqref{eq:eR}, and the matrix $E(R_e)$, defined in \eqref{eq:E}, which appears in the time derivative of the attitude error vector, namely, $\dot{e}_R=E(R_e)\omega_e$. More precisely, the metric $\Psi=\text{tr}[I-R_e]$ constitutes the \textit{quadratic} term for the angular position in the candidate Lyapunov function \eqref{eq:V}, and its left-trivialized derivative $e_R=0.5(R_e-R_e^{\top})^{\vee}$ constitutes the \textit{proportional} control part of the dynamic compensator \eqref{eq:Controller}. In addition, the error vector $e_R$ is also used to form cross-terms with $(\omega_e,x_K)$ in \eqref{eq:V}, as a result of which the matrix $E(R_e)$ appears in the Lyapunov rate \eqref{eq:Vdot}. This shows the crucial role that the choice of metric (or configuration error function) plays in the analytical framework, and also suggests a way forward for extending the framework to other commonly used metrics on $\mathrm{SO(3)}$. As an example, we consider the attitude error function proposed in \cite{lee2011geometric,lee2012exponential}. This error function was used to develop an almost globally exponentially stabilizing PD-type attitude control law which, when compared to control laws based on the chordal metric, yields better tracking performance especially in the case of rotational errors close to $\pi$ radians. 

The attitude error function, proposed in \cite{lee2011geometric,lee2012exponential}, is defined as follows:
\begin{equation}
\Psi_{q}(R_e)\isdef2-\sqrt{1+\text{tr}[R_e]}.
\end{equation}
The corresponding attitude error vector is given by:
\begin{equation}
e_{q}\isdef\frac{1}{2\sqrt{1+\text{tr}[R_e]}}(R_e-R_e^\top)^{\vee}.
\end{equation}
From \cite[Proposition 2]{lee2011geometric}, we know that:
\begin{equation}
\dot{\Psi}_{q}=e_{q}\cdot\omega_e.\label{eq:Psi_dot}
\end{equation}
Similarly, using \cite[Proposition 1]{lee2011geometric}, we observe that in the sublevel set
\begin{equation}
L_2\isdef\{R_e\in \mathrm{SO(3)}:\Psi_{q}(R_e)<2\},
\end{equation}
$\Psi_q$ is locally quadratic, that is,
\begin{equation}
||e_{q}||^2\leq\Psi_q(R_e)\leq2||e_{q}||^2.\label{eq:Psiq_LBUB}
\end{equation}
Following the same steps as in \cite[Proposition 2]{lee2011geometric}, and using the axis-angle representation \eqref{eq:Re_Axang}, we obtain the rate of change of the attitude error vector as:
\begin{align*}
\dot{e}_{q} & =\frac{1}{2\sqrt{1+\text{tr}[R_e]}}(\text{tr}[R_e]I-R_e^{\top}+2e_{q}e_{q}^{\top})\omega_e \\
& =\frac{1}{2\sqrt{2+2\cos\theta_e}}\left[(1+2\cos\theta_e)I-I+(\sin\theta_e)\alpha^{\times}\right.\\
&\quad\;\left.-(1-\cos\theta_e)\alpha^{\times}\alpha^{\times}+\frac{2\sin^2\theta_e}{2+2\cos\theta_e}\alpha\alpha^{\top}\right]\omega_e\\
& =\frac{1}{2\sqrt{2+2\cos\theta_e}}\left[(2\cos\theta_e)I+(\sin\theta_e)\alpha^{\times}\right.\\
&\quad\;\left.-(1-\cos\theta_e)(\alpha\alpha^{\top}-I)+(1-\cos\theta_e)\alpha\alpha^{\top}\right]\omega_e.
\end{align*}
Consequently, we can express $\dot{e}_q$ as:
\begin{equation}
\dot{e}_q=\frac{1}{2}E_q(\alpha,\theta_e)\omega_e,
\end{equation}
where
\begin{equation}
E_q(\alpha,\theta_e)\isdef\frac{1}{\sqrt{2+2\cos\theta_e}}[(1+\cos\theta_e)I+(\sin\theta_e)\alpha^{\times}].
\end{equation}
Using the MATLAB Symbolic Computation Tool, we observe that the eigenvalues of the matrix $E_q^{\top}E_q$ are $1$, $1$, and $\frac{1}{2}(1+\cos\theta_e)$. From this, it follows that $E_q^{\top}E_q\leq I$.

Next, we modify the dynamic compensator \eqref{eq:Controller} by replacing $e_{R}$ by $e_{q}$. More precisely, we use the control law \eqref{eq:tau}-\eqref{eq:uC} together with the following dynamic compensator:
\begin{equation}
\begin{split}\dot{x}_{K} & =A_{K}x_{K}+B_{\theta}e_{q}+B_{\omega}\omega_{e},\\
u & =C_{K}x_{K}+D_{\theta}e_{q}+D_{\omega}\omega_{e}.
\end{split}
\label{eq:Controller_q}
\end{equation}
The corresponding closed-loop error system is:
\begin{equation}
\begin{split}\dot{R}_{e} & =R_{e}\omega_{e}^{\times},\\
J\dot{\omega}_{e} & =C_{K}x_{K}+D_{\theta}e_{q}+D_{\omega}\omega_{e},\\
\dot{x}_{K} & =A_{K}x_{K}+B_{\theta}e_{q}+B_{\omega}\omega_{e}.
\end{split}
\label{eq:ClosedLoop_q}
\end{equation}
We consider the following candidate Lyapunov function:
\begin{equation}
\begin{split}V_{q}\isdef & \;2p_{11}\Psi_{q}(R_e)+\omega_{e}\cdot P_{22}J\omega_{e}+2e_{q}\cdot P_{21}^{\top}J\omega_{e}\\
 & \quad\;+x_{K}\cdot P_{33}x_{K}+2x_{K}\cdot P_{31}e_{q}+2x_{K}\cdot P_{32}J\omega_{e}.
\end{split}
\label{eq:V2}
\end{equation}
Using the lower bound in \eqref{eq:Psiq_LBUB}, we obtain the following sufficient condition for $V_{q}$ to be positive definite:
\begin{equation}
\mathcal{P}_q\isdef\begin{bmatrix}2p_{11}I & * & *\\
JP_{21} & P_{22}J & *\\
P_{31} & P_{32}J & P_{33}
\end{bmatrix}>0.\label{eq:V2_PosDef}
\end{equation}
Carrying out the same steps as in the proof of Proposition 1, we can express the Lyapunov rate $\dot{V}_{q}$ along trajectories of \eqref{eq:ClosedLoop_q} as follows:
\begin{equation}
\begin{split}
\dot{V}_{q} & =x^{\top}\mathcal{M}_0x+\omega_{e}\cdot JP_{21}E_{q}(\alpha,\theta_e)\omega_{e}\\
 & \quad\;+x_{K}\cdot P_{31}E_{q}(\alpha,\theta_e)\omega_{e},
\end{split}\label{eq:V2dot}
\end{equation}
where $x=\text{col}(e_{q},\omega_e,x_{K})$. Applying the variable elimination lemma, we observe that the Lyapunov rate \eqref{eq:V2dot} is negative definite in the sublevel set $L_2$ if and only if there exist positive scalars $(\tau_1,\tau_2)$ such that
\begin{equation}
\begin{split}
&x^{\top}\mathcal{M}_0x+\frac{1}{4\tau_1}x_{K}^{\top}(P_{31}P_{31}^{\top})x_{K}+\tau_1\omega_{e}^{\top}\omega_{e}\\
&\quad\;+\frac{1}{4\tau_2}\omega_{e}^{\top}J(P_{21}P_{21}^{\top})J\omega_{e}+\tau_2\omega_{e}^{\top}\omega_{e}<0,
\end{split}\label{eq:V2dot_A}
\end{equation}
where the matrix $\mathcal{M}_0$ and its submatrices are defined in Proposition \ref{prop:Vdot}. Using similar steps as carried out in Section \ref{sec:LyapunovAnalysis_Chordal} for the chordal metric, we obtain the following sufficient condition for the negative definiteness of the Lyapunov rate $\dot{V}_q$:
\begin{equation}
\begin{split}
& \mathcal{M}_3\isdef\begin{bmatrix}M_{11} & * & *\\
M_{21} & M_{22}+(\tau_1+\tau_2)I + N_2 & *\\
M_{31} & M_{32} & M_{33}+N_3
\end{bmatrix} <0,\\
&\quad\quad\quad \begin{bmatrix}N_2 & JP_{21}\\P_{21}^{\top}J & 4\tau_2 I\end{bmatrix}\geq0,\quad\quad\quad \begin{bmatrix}N_3 & P_{31}\\P_{31}^{\top} & 4\tau_1 I\end{bmatrix}\geq0.
\end{split}\label{eq:V2dot_NegDef_LMI}
\end{equation}
\begin{corollary}
Consider the closed-loop tracking error system \eqref{eq:ClosedLoop_q} and the candidate Lyapunov function \eqref{eq:V2}. Suppose that the Lyapunov function coefficients satisfy the linear matrix inequalities \eqref{eq:V2_PosDef} and \eqref{eq:V2dot_NegDef_LMI}, with the relevant submatrices defined in \eqref{eq:Mij}. Then, $V_{q}>0$, $\dot{V}_{q}<0$, and the controller \eqref{eq:Controller_q} almost globally asymptotically stabilizes the desired equilibrium $(R_e,\omega_e,x_K)=(I,0,0)$ of the tracking error system \eqref{eq:ClosedLoop_q}.
\end{corollary}

\begin{remark}
Similar to \cite{lee2011geometric,lee2012exponential}, it can be shown that the closed-loop tracking error system \eqref{eq:ClosedLoop_q} is almost globally exponentially stable (AGES). Likewise, using similar arguments as in \cite[Proposition 2]{lee2015global}, it can be shown that the closed-loop tracking error system \eqref{eq:ClosedLoop} is almost semi-globally exponentially stable (AsGES).
\end{remark}

\section{Case Study}\label{sec:SimulationStudy}

In this section, we demonstrate how the theoretical framework developed above can be used to design and analyze higher-order geometric nonlinear attitude controllers of the form \eqref{eq:Controller}. As our case study, we consider the attitude control problem for a multicopter UAV. A detailed description of the UAV model is available in \cite{khan2020robust}. Here, we assume that its inertia matrix is
\begin{equation}
J = \begin{bmatrix}0.0411&0.002&-0.001\\0.002&0.0478&0.003\\-0.001&0.003&0.0599\end{bmatrix}\;\text{kg}/\text{m}^2.\label{eq:J}
\end{equation}
\textcolor{black}{We note that the inertia matrix used in \cite{khan2020robust} is diagonal. Here, we have included off-diagonal entries merely to demonstrate that the proposed analytical framework can be used with non-diagonal inertia matrices as well.}

The case study is divided into two parts. In the first part, we design different PID-type attitude control laws for the multicopter UAV model. In particular, we consider the geometric nonlinear PID controller proposed in \cite{goodarzi2013geometric}, as well as cascade P/PI and P/PID controllers.
The P/PID controller includes an additional gain which acts on the \emph{filtered} derivative of the angular velocity.
In the second part of the case study, we describe how the LMI-based analytical framework proposed in Corollary \ref{Theorem:Chordal_LMI} can be used in conjunction with standard linearization-based design methods to obtain higher-order geometric nonlinear controllers with local performance and almost global asymptotic stability guarantees.

\subsection{PID-based Geometric Attitude Control}\label{sec:PID}


\subsubsection{Baseline PID Controller}

The first controller under consideration is the PID controller proposed in \cite{goodarzi2013geometric}. The state-feedback component of this controller can be expressed as:
\begin{equation}
\begin{split}
u&=-k_Pe_R-k_D\omega_e-k_Ie_I,\\
\dot{e}_{I}&=ce_R+\omega_e.
\end{split}
\end{equation}
where $(k_P,k_D,k_I,c)\in\Rset$ are positive constants. A state-space realization is given by:
\begin{equation}
\left[\begin{array}{c|c|c}
             A_K & B_{\theta} & B_{\omega} \\ \hline
             C_K & D_{\theta} & D_{\omega}
           \end{array}\right]
= \left[\begin{array}{c|c|c}
             0 & cI & I \\ \hline
             -k_I I & -k_P I & -k_D I
           \end{array}\right].\nonumber
\end{equation}
We use decoupled single-axis linearized models to tune the controller gains for the roll, pitch, and yaw axes. In particular, we choose the controller gains as
\begin{equation}
(k_P,k_D,k_I,c) = (7.3878,1.7238,0.9358,5).\label{eq:Gains_BaselinePID}
\end{equation}
Figure 1 plots the loopshape for the baseline PID controller. This design is used as a benchmark for tuning the cascade P/PI and P/PID controllers. More precisely, the controllers are tuned such that their loop crossover frequencies and closed-loop bandwidths are similar. In particular, the crossover frequencies lie in the interval $29-42\;\text{rad/s}$, and the closed-loop bandwidth is approximately $5\;\text{rad/s}$.
%
Lastly, using the LMIs in Corollary \ref{Theorem:Chordal_LMI}, we confirm that the gains in \eqref{eq:Gains_BaselinePID} almost globally asymptotically stabilize the desired equilibrium of the nonlinear tracking error system \eqref{eq:ClosedLoop}.


\begin{figure}
\begin{center}
\includegraphics[width=0.9\linewidth]{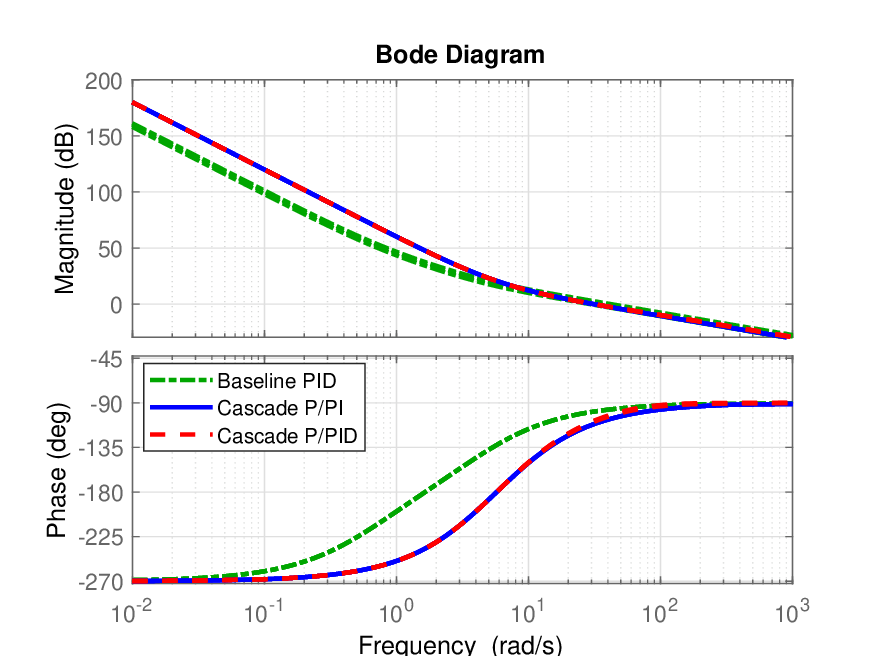}
\caption{Loopshapes for the three PID controllers under consideration: baseline PID (blue, solid); cascade P/PI (red, dashed dot); cascade P/PID (green, dashed). The three controllers are tuned such that, for each axis, the loop crossover frequencies and closed-loop bandwidths are similar.}\label{fig:10}
\end{center}
\end{figure}





\subsubsection{Cascade P/PI Controller}

Next, we consider a cascade P/PI controller with PI compensation in the (inner) rate loop and proportional control in the (outer) attitude loop. In particular, the control law is given by
\begin{equation}
u=K_{\omega}e_{\omega}+K_Ie_I,\quad\dot{e}_{I}=e_{\omega},\label{eq:u_PPI}
\end{equation}
where $e_{\omega}=\omega_{\text{ref}}-\omega$ is the tracking error for the rate controller, $\omega_{\text{ref}}=-K_Re_R+R_e^{\top}\omega_d$ is the velocity reference generated by the attitude controller, and $(K_R,K_{\omega},K_I)\in\Rset^{3\times3}$ are symmetric positive definite gain matrices. We choose these gains as
\begin{equation}
(K_R,K_{\omega},K_I) = (4.383I,2\omega_n J,\omega_n^2 J),\label{eq:Gains_PPI}
\end{equation}
where $\omega_n=15\;\text{rad/s}$. 
We note that this tuning approach is motivated by the tuning methodology followed in \cite[Section 5]{invernizzi2022global}. For the inner loop tracking error, we observe that:
\[
e_{\omega}=-K_Re_R-(\omega-R_e^{\top}\omega_d)=-K_Re_R-\omega_e,
\]
where $\omega_e$ is the attitude error defined in \eqref{eq:Errors}. Consequently, the cascade P/PI controller \eqref{eq:u_PPI} is re-stated as:
\begin{equation}
\begin{split}
u&=-K_{\omega}K_Re_R-K_{\omega}\omega_e+K_Ie_I,\\
\dot{e}_{I}&=-K_Re_R-\omega_e.
\end{split}\label{eq:u_PPI_Restated}
\end{equation}
A state space realization is given by:
\begin{equation}
\left[\begin{array}{c|c|c}
             A_K & B_{\theta} & B_{\omega} \\ \hline
             C_K & D_{\theta} & D_{\omega}
           \end{array}\right]
= \left[\begin{array}{c|c|c}
             0 & K_R & I \\ \hline
             -K_I & -K_{\omega}K_R & -K_\omega
           \end{array}\right].\nonumber
\end{equation}
Using the gains \eqref{eq:Gains_PPI} and the above state space realization, we confirm that the LMIs in Corollary \ref{Theorem:Chordal_LMI} are feasible, and that the cascade P/PI controller almost globally asymptotically stabilizes the closed-loop tracking error system \eqref{eq:ClosedLoop}. 
%


\subsubsection{Cascade P/PID Controller}

For our third and final PID variant, we consider the combination of cascade P/PI control \eqref{eq:u_PPI_Restated} and acceleration feedback using a filtered derivative of the angular velocity. In particular, we obtain the filtered angular acceleration $\sigma$ as
\begin{equation}
\dot{q}=-Nq-N\omega,\quad\sigma=Nq+N\omega,\label{eq:omegadot}
\end{equation}
where $N\in\Rset^{3\times3}$ is a diagonal positive definite gain matrix. Next, we design the following cascade P/PID controller:
\begin{equation}
\begin{split}
u&=-K_{\omega}K_Re_R-K_\omega\omega_e+K_Ie_I-K_A\sigma,\\
\dot{e}_{I}&=-K_Re_R-\omega_e,
\end{split}\label{eq:u_PPID}
\end{equation}
where the gain $K_A$ acts on the filtered angular acceleration. We recall that for the attitude regulation and setpoint tracking problems, the desired angular velocity $\omega_d=0$ and the velocity error $\omega_e=\omega$. \textcolor{black}{Consequently, when the controller \eqref{eq:omegadot}-\eqref{eq:u_PPID} is used for attitude regulation or setpoint tracking, a state space realization can be obtained as:}
\begin{equation}
\left[\begin{array}{cc|c|c}
0 &  0 & K_R & I\\
0 & -N & 0 & -N\\ \hline
-K_I & -K_AN & -K_\omega K_R & -(K_\omega + K_AN)
\end{array}\right].
\end{equation}
In addition to the cascade P/PI gains defined in \eqref{eq:Gains_PPI}, we choose the following controller gains:
\begin{align*}
K_A &= 0.00263 I, &N &= 75 I.
\end{align*}
We verify that these gains satisfy the LMI-based stabilization conditions given in Corollary \ref{Theorem:Chordal_LMI}, and thus ensure that the closed-loop attitude control system (for attitude regulation and setpoint tracking) is AGAS. 




\subsection{Higher-Order Geometric Attitude Control}

In the second part of our case study, we show how the proposed analytical framework can be used as part of a two-step design method for obtaining higher-order geometric nonlinear controllers for the attitude control problem. In the first step, we linearize the attitude kinematics and dynamics around an operating point and design a robust dynamic compensator using standard tools from LTI systems theory. This step yields the controller matrices to be used in the higher-order geometric nonlinear (or \emph{geometrized} linear) controller, described in \eqref{eq:Controller}. Then, we check whether the stabilization conditions stipulated in Corollary \ref{Theorem:Chordal_LMI} hold. This step amounts to solving a feasibility problem in which the constraint set consists of linear matrix inequalities. The successful solution of this problem allows us to conclude that the higher-order geometric nonlinear controller, designed in the first step, almost globally asymptotically stabilizes the closed-loop tracking error system. Together, the two steps ensure that the designed controller yields adequate local performance and almost global asymptotic stability for the multicopter UAV under consideration.

\subsubsection{Controller Design}

In the first step, we use the linearized model
\begin{align*}
\dot{\xi} &=\omega,&J\dot{\omega} &=u,
\end{align*}
to design a linear controller capable of tracking a constant or slowly-varying reference angular trajectory $\xi_d$. To this end, we use decoupled single-axis linearized models to design individual controllers for the roll, pitch, and yaw axes. For each axis, we adopt a cascade architecture in which the (inner) rate loop uses a lead filter and the (outer) attitude loop uses a lead-lag filter (or implementable PID controller). In particular, the linear controller is given by
\begin{equation}
u_0=K_{\omega}(s)(\omega_{\text{ref}}-\omega),\quad\omega_{\text{ref}}=K_R(s)(\xi_d-\xi),\label{eq:u1_Lin}
\end{equation}
where $K_{\omega}$ and $K_R$ denote the inner and outer loop controllers, respectively. Choosing these controllers as
\begin{equation}
\begin{split}
K_\omega(s) &= 5\left(\frac{s + 2}{s + 2.5}\right)I_3,\\
K_R(s) &= 37.5\left(\frac{s+1.653}{s+2.5}\right) \left(\frac{s+0.05042}{s + 0.01}\right)I_3,
\end{split}\label{eq:LeadLag}
\end{equation}
we obtain a closed-loop system with \textcolor{black}{adequate gain and phase margins for each axis, as well as acceptable performance characteristics in terms of rise time, settling time, and percentage overshoot. The second column of Table \ref{tab:tab1} lists the performance characteristics for the controller defined in \eqref{eq:u1_Lin}-\eqref{eq:LeadLag}. The last entry, namely $\gamma$, indicates the controller's ability to handle uncertainties at the plant input. In particular, this number indicates that for each decoupled single-axis linearized plant, uncertainties at the plant input will not be amplified beyond $\gamma=1.93348$. Moreover, a reduction in the value of $\gamma$ will improve the control system's robustness to input uncertainties.} 

\ra{1.3}
\begin{table}[h]
\begin{center}
\caption{Closed-Loop Performance Characteristics for the Initial Design and the Robust Design}
\begin{tabular}{ccc}
\hline
Property & Initial Design & Robust Design \\ \cmidrule(lr){1-3}
Disk Gain Margin   & $\pm12.13$ dB      & $\pm16.36$ dB \\
Disk Phase Margin  & $62.19$ deg        &  $72.71$ deg \\
Rise Time (ms)     & $[45.0,43.6,42.3]$ & $[49.3,48.4,47.8]$ \\
Settling Time (ms) & $[88.4,73.7,64.7]$ & $[89.2,82.3,76.2]$ \\
Overshoot (\%)     & $[ 0.0, 0.0, 0.9]$ & $[ 0.0, 0.0, 0.3]$ \\
$\gamma$      & 1.93348 & 1.68569  \\ \hline
\end{tabular}
\label{tab:tab1}
\end{center}
\end{table}
\ra{1.0}

\begin{figure}
\begin{center}
\includegraphics[width=0.9\linewidth]{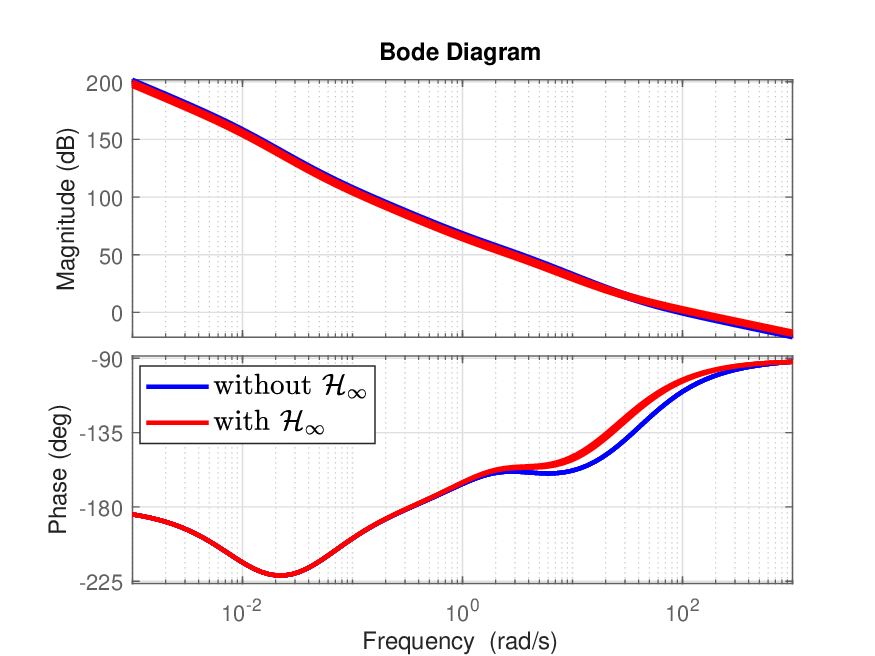}
\caption{Loopshapes for the initial and final designs: initial design (blue); final design (red). The final design uses $H_{\infty}$ loopshaping to improve the robustness of the initial design to uncertainties at the plant input.}\label{fig:1}
\end{center}
\end{figure}

\begin{figure}
\begin{center}
\includegraphics[width=0.9\linewidth]{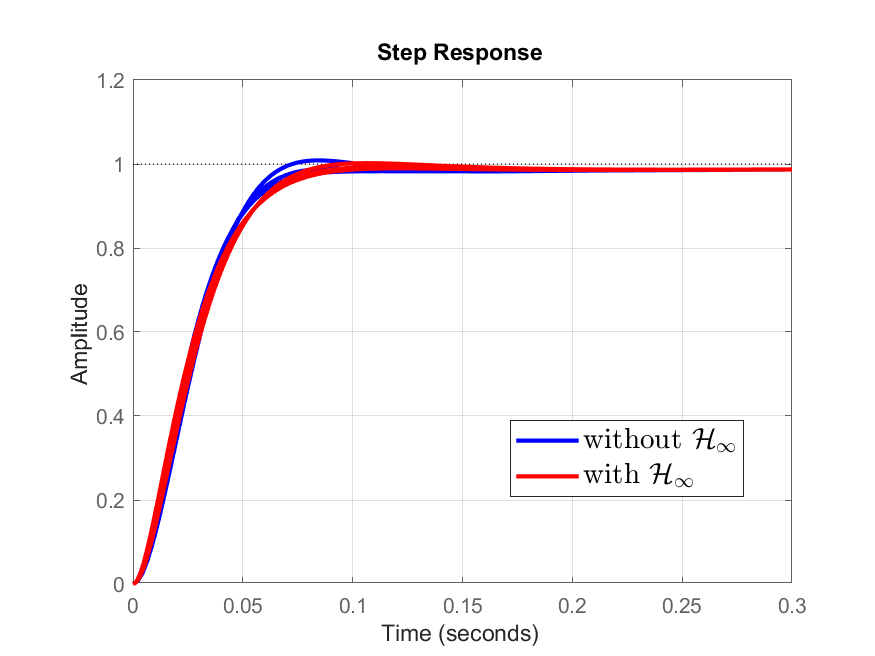}
\caption{Step responses, for each axis, obtained using the initial and final designs: initial design (blue); final design (red). The final design uses $H_{\infty}$ loopshaping to improve the robustness of the initial design to uncertainties at the plant input.}\label{fig:2}
\end{center}
\end{figure}

To that end, we use the Glover-McFarlane method \cite{mcfarlane1990robust,mcfarlane1992loop} to design a robust controller for the inner loop of each axis. More precisely, after opening the loop at the plant input $u$, we use MATLAB's `ncfsyn' command to compute a normalized coprime factor $\mathcal{H}_{\infty}$ loopshaping controller. Appending this controller to the inner loop, we obtain the final linear controller as
\begin{equation}
u_{\text{lin}} = -K_\infty(s)K_\omega(s)(\omega_{\text{ref}}-\omega),\label{eq:u2_Lin}
\end{equation}
where $K_\infty$ denotes the  $\mathcal{H}_{\infty}$ loopshaping controller. Figures \ref{fig:1} and \ref{fig:2}
depict, respectively, the loopshapes and closed-loop step responses for the initial \eqref{eq:LeadLag} and final designs \eqref{eq:u2_Lin}. The loopshapes and step responses are very similar. However, the final design $u_{\text{lin}}$ improves the value of $\gamma$ from $1.93348$ to $1.68569$. This represents an improvement of approximately $15\%$ in the controller's ability to handle uncertainties at the plant input. However, this step also increases the controller order from $3$ to $7$ for each axis. As a result, the complete controller has $21$ states.

A minimal state-space realization for the final design \eqref{eq:u2_Lin} is given by:
\begin{equation}
\begin{split}\dot{x}_{K} & =A_{K}x_{K}+B_{\theta}(\xi-\xi_d)+B_{\omega}\omega,\\
u_{\text{lin}} & =C_{K}x_{K}+D_{\theta}(\xi-\xi_d)+D_{\omega}\omega.
\end{split}\label{eq:Controller_Linear}
\end{equation}
Replacing the terms $(\xi-\xi_d,\omega)$ by the attitude and velocity errors $(e_R,\omega_e)$, defined in \eqref{eq:Errors} and \eqref{eq:eR}, we obtain the corresponding geometric nonlinear controller:
\begin{equation}
\begin{split}\dot{x}_{K} & =A_{K}x_{K}+B_{\theta}e_{R}+B_{\omega}\omega_{e},\\
u & =C_{K}x_{K}+D_{\theta}e_{R}+D_{\omega}\omega_{e}.
\end{split}\label{eq:Controller_Geom}
\end{equation}
For this controller, we confirm that the stabilization conditions given in Corollary \ref{Theorem:Chordal_LMI} are satisfied. \textcolor{black}{In particular, we confirm these conditions by formulating and solving a semidefinite programming problem using MATLAB R2020a, and the YALMIP and SeDuMi toolboxes \cite{lofberg2004yalmip,sturm1999using}}. Consequently, we conclude that the geometric nonlinear controller \eqref{eq:Controller_Geom} almost globally asymptotically stabilizes the desired equilibrium of the closed-loop tracking error system.

\subsubsection{Simulation Results}

\begin{figure}
\begin{center}
\includegraphics[width=0.9\linewidth]{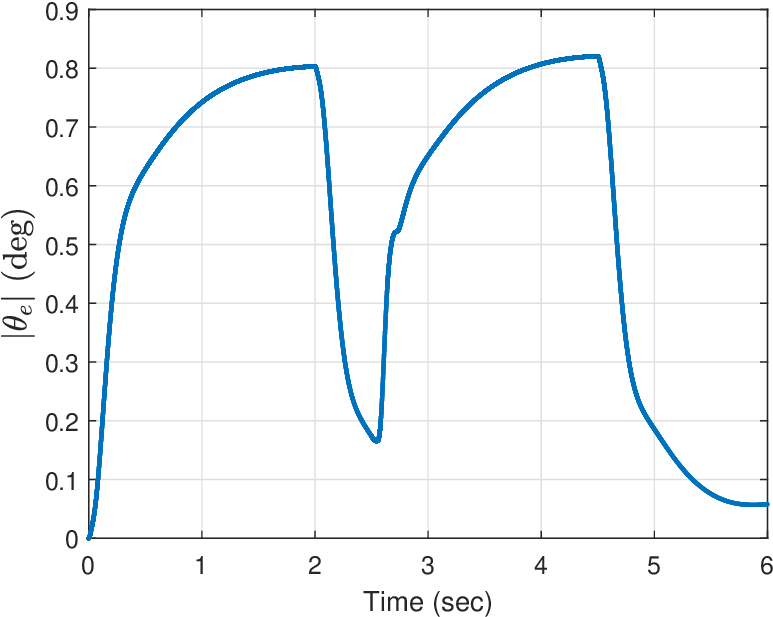}
\caption{Attitude error during agile reference tracking. The attitude error $\theta_e$ corresponds to the angle of rotation associated with the attitude error $R_e=R_d^{\top}R\in\mathrm{SO(3)}$.}\label{fig:4}
\end{center}
\end{figure}
\begin{figure}
\begin{center}
\includegraphics[width=0.9\linewidth]{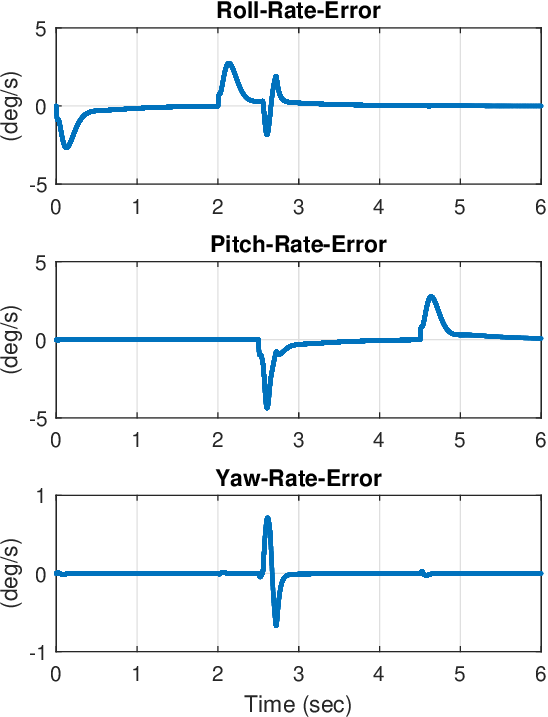}
\caption{Roll, pitch and yaw rate errors during agile reference tracking: roll-rate error (top); pitch-rate error (center); yaw-rate error (bottom).}\label{fig:5}
\end{center}
\end{figure}

In this section, we present simulation results demonstrating the tracking performance of the higher-order geometric nonlinear controller designed above. We assume that the hexacopter is initially at rest with initial attitude $R(0)=I$. The commanded trajectory consists of two flips about the roll axis followed by two flips about the pitch axis before returning to the initial attitude. More precisely, the following maneuver is commanded:
\begin{equation}\label{eq:Ref_Traj}
       \bar{R}_d(t) = \begin{cases}
        \exp(2\pi t e_1^{\times}), & \mbox{if } 0 \leq t \leq 2 \\
        \exp(2\pi(t-2.5) e^\times_2), & \mbox{if } 2.5 < t \leq 4.5 \\
        I, & \mbox{otherwise},
      \end{cases}
\end{equation}
where $e_1 = [1,0,0]^{\top}$ and $e_2 = [0,1,0]^{\top}$. We execute this maneuver by generating a filtered reference $(R_d,\omega_d)$ using the second-order geometric filter described in \cite[Section VI-C]{invernizzi2020robust}. In particular, we design this filter so that its linearized counterpart has a natural frequency of $15\;\text{rad}/\text{s}$ and a damping ratio of $0.707$. For numerical simulations, we employ a high-fidelity nonlinear model which includes: $\left.1\right)$ actuator dynamics and saturation limits on the RPMs for each motor; and $\left.2\right)$ a pseudo-inverse-based control allocation scheme for distributing the desired torques among the motors. More details on the simulation model are available in \cite{khan2020robust}. We plot the results in terms of the angle of rotation associated with the attitude tracking error $R_e=R_d^{\top}R$. From Figs. \ref{fig:4} and \ref{fig:5}, we see that the attitude error remains below $1$ degree during the maneuver, and the velocity errors subside quickly.


\section{Conclusion}\label{sec:Conclusion}

This paper presented an analytical framework which can be used to analyze the closed-loop stability properties of higher-order geometric nonlinear control laws for attitude control on the Special Orthogonal Group $\mathrm{SO(3)}$. Using Lyapunov stability theory, the proposed analytical framework extended existing results on the almost global asymptotic stability (AGAS) properties of geometric nonlinear PD and PID controllers. The Lyapunov approach followed in the present work was inspired by quadratic Lyapunov functions of the form used in the stability analysis of LTI systems. In particular, a sufficient condition was obtained which ensured the positive definiteness of the candidate Lyapunov function, and a necessary and sufficient condition which ensured the negative definiteness of the Lyapunov rate along trajectories of the closed-loop tracking error system. These conditions were expressed in the form of matrix inequalities, which were then relaxed to linear matrix inequalities (LMIs). The proposed LMIs can be used to search for appropriate Lyapunov function coefficients certifying the AGAS properties of a given higher-order geometric nonlinear control law. The applicability of the approach to practical problems was demonstrated by designing and analyzing a $21$-state geometric nonlinear control law for agile attitude tracking by a multicopter. In future work, we hope to extend the approach to the problem of controller design using, for instance, bilinear matrix inequalities (BMIs) and/or alternating semidefinite programming (SDP). We hope to also include actuator dynamics in the problem formulation, as well as to address issues related to imperfect cancellation torques due to, for instance, imprecise knowledge of the inertia matrix, measurement noise, etc.


\bibliographystyle{plain}        
\bibliography{Library}           

\end{document}